 \documentclass[10pt,preprint2]{aastex}

\newcommand{\Ell}{E_\parallel}      % E_\parallel
\newcommand{\rhoGJ}{\rho_{{\rm GJ}}}  % Goldreich-Julian charge density
\newcommand{\rlc}{\varpi_{\rm LC}} % light cylinder radius
\newcommand{\inc}{\alpha_{\rm i}}  % inclination angle of oblique rotator

\shorttitle{High-energy Emission from Pulsars}
\shortauthors{Hirotani}

\begin{document}

%% LaTeX will automatically break titles if they run longer than
%% one line. However, you may use \\ to force a line break if
%% you desire.

\title{High-energy Emission from Pulsar Outer Magnetospheres}

%% Use \author, \affil, and the \and command to format
%% author and affiliation information.
%% Note that \email has replaced the old \authoremail command
%% from AASTeX v4.0. You can use \email to mark an email address
%% anywhere in the paper, not just in the front matter.
%% As in the title, use \\ to force line breaks.

\author{Kouichi Hirotani}
\affil{ASIAA/National Tsing Hua University - TIARA,\\
       PO Box 23-141, Taipei, Taiwan\footnote{
            Postal address: 
            TIARA, Department of Physics, 
            National Tsing Hua University,
            101, Sec. 2, Kuang Fu Rd.,Hsinchu, Taiwan 300}
      }
\email{hirotani@tiara.sinica.edu.tw}

%% Notice that each of these authors has alternate affiliations, which
%% are identified by the \altaffilmark after each name.  Specify alternate
%% affiliation information with \altaffiltext, with one command per each
%% affiliation.

% \altaffiltext{1}{Visiting Astronomer, ...}
% \altaffiltext{2}{Society of Fellows, Harvard University.}
% \altaffiltext{3}{present address: Center for Astrophysics,
%     60 Garden Street, Cambridge, MA 02138}
% \altaffiltext{4}{Visiting Programmer, Space Telescope Science Institute}
% \altaffiltext{5}{Patron, Alonso's Bar and Grill}

%% Mark off your abstract in the ``abstract'' environment. In the manuscript
%% style, abstract will output a Received/Accepted line after the
%% title and affiliation information. No date will appear since the author
%% does not have this information. The dates will be filled in by the
%% editorial office after submission.

\begin{abstract}
We investigate particle accelerators in rotating neutron-star 
magnetospheres, by simultaneously solving 
the Poisson equation for the electrostatic potential
together with the Boltzmann equations for 
electrons, positrons and photons on the poloidal plane.
Applying the scheme to the three pulsars, Crab, Vela and PSR~B1951+32,
we demonstrate that the observed phase-averaged spectra are 
basically reproduced from infrared to very high energies.
It is found that the Vela's spectrum in 10-50 GeV
is sensitive to the three-dimensional magnetic field configuration
near the light cylinder; thus, 
a careful argument is required
to discriminate the inner-gap and outer-gap emissions using
a gamma-ray telescope like GLAST.
It is also found that PSR~B1951+32 
has a large inverse-Compton flux in TeV energies,
which is to be detected by ground-based air Cerenkov telescopes
as a pulsed emission.
\end{abstract}

%% Keywords should appear after the \end{abstract} command. The uncommented
%% example has been keyed in ApJ style. See the instructions to authors
%% for the journal to which you are submitting your paper to determine
%% what keyword punctuation is appropriate.

\keywords{gamma-rays: theory 
       -- magnetic fields 
       -- methods: numerical
       -- pulsars: individual(\objectname{B1951+32,Crab,Vela})}

%% From the front matter, we move on to the body of the paper.
%% In the first two sections, notice the use of the natbib \citep
%% and \citet commands to identify citations.  The citations are
%% tied to the reference list via symbolic KEYs. The KEY corresponds
%% to the KEY in the \bibitem in the reference list below. We have
%% chosen the first three characters of the first author's name plus
%% the last two numeral of the year of publication as our KEY for
%% each reference.

%% Authors who wish to have the most important objects in their paper
%% linked in the electronic edition to a data center may do so by tagging
%% their objects with \objectname{} or \object{}.  Each macro takes the
%% object name as its required argument. The optional, square-bracket 
%% argument should be used in cases where the data center identification
%% differs from what is to be printed in the paper.  The text appearing 
%% in curly braces is what will appear in print in the published paper. 
%% If the object name is recognized by the data centers, it will be linked
%% in the electronic edition to the object data available at the data centers  
%%
%% Note that for sources with brackets in their names, e.g. [WEG2004] 14h-090,
%% the brackets must be escaped with backslashes when used in the first
%% square-bracket argument, for instance, \object[\[WEG2004\] 14h-090]{90}).
%%  Otherwise, LaTeX will issue an error. 

\section{Introduction}
\label{sec:intro}
Pulsars form the second most numerous class of objects
detected in high-energy $\gamma$-rays.
To date, six have been detected by the 
Energetic Gamma Ray Experiment Telescope (EGRET) 
aboard the Compton Gamma Ray Observatory. 
% The success of EGRET in the 30~MeV--30~GeV energy range
% suggests a candidate source list for follow-up observations
% in the VHE range, i.e., $h\nu>100$~GeV.
% VHE emission has been deteced from the direction of the 
% Crab nebula (Vacanti 1991),
% the Vela pulsar (Takanori 1996), and
% PSR~B1706--44 (Kifune et al. 1995).
% However, no evidence has been found for periodic emission in VHE;
% only upper limits have been obtained for the six pulsars
% (Thompson 1997; for Crab, see Lessard et al. 1999;
%  for Vela, Yoshikoshi et al. 1997;
%  for B1951+32, Srinivasan et al. 1997).
The $\gamma$-ray pulsations observed from these objects
are particularly important as a direct signature of
non-thermal processes in rotating neutron-star magnetospheres,
and potentially should help to discriminate among different emission models.

The pulsar magnetosphere can be divided into two zones: 
The closed zone filled with a dense plasma corotating with the star,
and the open zone in which plasma flows along the open field lines
to escape through the light cylinder.
The last-open field lines form the border of the open magnetic field
line bundle.
In all the pulsar emission models, 
particle acceleration takes place in this open zone.
In inner-gap (IG) models,
which adopts particle acceleration within several neutron-star
radii above the polar-cap surface,
the energetics and pair cascade spectrum have had success
in reproducing the observations
(e.g., Daugherty \& Harding 1982, 1996).
However, the predicted beam size 
is too small to produce the wide pulse profiles that are observed
from high-energy pulsars.
Seeking the possibility of a wide hollow cone emission
due to flaring field lines,
Muslimov and Harding (2004) extended the idea
of Arons (1983) and proposed a slot-gap (SG) model, 
in which emission takes place very close to the last-open field line
from the stellar surface to the outer magnetosphere.
Since the SG model is an extension of the IG model 
into the outer magnetosphere,
a negative magnetic-field-aligned electric field, $\Ell$,
arises if the magnetic moment vector
points in the same hemisphere as the rotation vector.
However, the electric current induced by the negative $\Ell$
does not have a self-consistent closure within the model
(Hirotani~2006, hereafter H06).

% On these grounds, we are motivated by the need to contrive 
% an accelerator model that predicts a consistent current direction
% with the global requirement. 
To contrive an accelerator model that predicts an outward current 
in the lower latitudes (within the open zone),
it is straightforward to extend outer-gap (OG) models
(Cheng et al. 1986; Romani \& Yadigaroglu 1995, hereafter RY95;
 Cheng et al. 2000)
% which predict opposite $\Ell$ to IG models,
into the inner magnetosphere.
Extending several OG models
(Hirotani et al.~2003; Takata et al.~2004), 
H06 demonstrated that
the gap extends from the stellar surface to the outer magnetosphere,
that the positive $\Ell$ 
extracts ions from the star as a space-charge-limited flow (SCLF), 
% in the same manner as electrons are extracted in IG and SG models,
and that most photon emission takes place in the outer magnetosphere
because $\Ell$ is highly screened inside the null surface
owing to the discharge of the created pairs.
In the present letter,
we formulate the scheme (Beskin et al.~1992) in \S~\ref{sec:basic_eqs},
apply it to three rotation-powered pulsars in \S~\ref{sec:appl},
and give a brief discussion in \S~\ref{sec:summary}.

\section{Gap Electrodynamics}
\label{sec:basic_eqs}
We follow the scheme described in \S~2 of H06
to solve the set of Maxwell and Boltzmann equations.
The first equation we have to consider is the Poisson equation
for the electro-static potential, $\Psi$.
As space is limited, we present its Newtonian expression,
$ -\nabla^2 \Psi = 4 \pi (\rho-\rhoGJ)$.
If the real charge density, $\rho$, deviates from the 
Goldreich-Julian (GJ) charge density,
$\rhoGJ$, in some region,
an acceleration electric field $\Ell \equiv -\partial \Psi/\partial s$
arises, 
where $s$ designates the distance along a magnetic field line.

The second equation we have to consider is the Boltzmann equations
for $e^\pm$'s.
Assuming a steady state in the frame of reference corotating with the
magnetosphere, we obtain 
\begin{equation}
  c\cos\chi \frac{\partial n_\pm}{\partial s}
  +\frac{dp}   {dt}\frac{\partial n_\pm}{\partial p}
  +\frac{d\chi}{dt}\frac{\partial n_\pm}{\partial \chi}
  = S_\pm,
 \label{eq:BASIC_2}
\end{equation}
where $c$ denotes the speed of light,
$n_+$ (or $n_-$) the dimensionless 
positronic (or electronic) distribution function 
normalized by the local GJ number density.
The temporal derivatives of momentum and pitch angle,
$dp/dt$ and $d\chi/dt$, and the collision term $S_\pm$
are explicitly given in H06.
Synchro-curvature radiation-reaction force is included as an external
force in $dp/dt$,
while the effects of inverse-Compton scatterings (ICS)
and (one-photon and two-photon) pair creation processes
are in $S_\pm$.

The third equation we have to consider is the Boltzmann equations
for photons.
Assuming a steady state, and neglecting azimuthal propagations, we obtain 
\begin{equation}
  c\frac{k^r}{\vert\mbox{\boldmath$k$}\vert}
   \frac{\partial g}{\partial r}
 +c\frac{k^\theta}{\vert\mbox{\boldmath$k$}\vert}
   \frac{\partial g}{\partial \theta}
 = S_\gamma(r,\theta,c\vert\mbox{\boldmath$k$}\vert,k^r,k^\theta),
  \label{eq:BASIC_3}
\end{equation}
where the wave numbers $k^r$ and $k^\theta$ are given by the ray path,
($r$,$\theta$) are the polar coordinates,
and the dimensionless photon distribution function $g$ is
normalized by the GJ number density at the stellar surface. 
ICS, synchro-curvature emission,
and the absorption are contained in $S_\gamma$.
In H06, the rate of synchrotron emission by secondary pairs
created outside the gap, was calculated
assuming a constant pitch angle.
However, it turns out that only 17~\% of the initial particle energy
is converted into radiation.
In this letter, we corrected this problem
by computing the pitch angle evolution of radiating particles,
which increases the synchrotron cooling time and hence 
recovers the time-integrated, radiated energy to 100~\% of
the initial particle energy.
Note that the gap electrodynamics investigated in H06 remains correct,
despite the insufficient secondary synchrotron fluxes in H06.

To solve the Poisson equation,
we impose $\Psi=0$ at the lower, upper, and inner ($s=0$) boundaries,
and $\Ell=-\partial\Psi/\partial s=0$ at the outer boundary.
Ion extraction rate is regulated by the condition $\Ell=0$ at $s=0$.
% To solve equations~(\ref{eq:BASIC_2}) and (\ref{eq:BASIC_3}),
% we assume that neither $e^\pm$'s nor photons are injected
% across the inner and outer boundaries into the gap.
We parameterize the trans-field thickness of the gap with $h_{\rm m}$.
If $h_{\rm m}=1$ the gap exists along all the open field lines,
while if $h_{\rm m}\ll 1$ the gap becomes transversely thin.

\section{Application to Individual Pulsars}
\label{sec:appl}
We apply the theory to three $\gamma$-ray pulsars,
Crab, Vela, and B1951+32,
focusing on their photon spectra.
Even near and outside the light cylinder,
photon emission and absorption occur effectively;
thus, equations~(\ref{eq:BASIC_2}) and (\ref{eq:BASIC_3})
are solved in $0<s<16\rlc$,
where $\Ell=0$ in $\varpi>0.9\rlc$;
$\rlc \equiv c/\Omega$ denotes the light cylinder radius, 
$\Omega$ the stellar rotation frequency,
and $\varpi$ the distance from the rotation axis.
The field line geometry in $0.9\rlc<\varpi<2\rlc$
mimics the aligned dipole in the force-free magnetosphere
(Contopoulos et al. 1999; Gruzinov 2005).
% Since time-dependent force-free simulations of oblique rotators
% have just begun,
% we do not adopt the field line configuration
% proposed by Spitkovsky (2006) 
% in this letter.
In $\varpi>2\rlc$,
the field lines are assumed to be straight 
so that they connect smoothly at $\varpi=2\rlc$.
% When we extend present analysis into three-dimensional magnetosphere
% and give more precise predictions,
% we believe it will be crucially important to adopt a correct 
% field line configuration inside and outside the light cylinder.
% However, to inquire further into the matter would lead us into
% a specialized area of force-free electrodynamics,
% and would undoubtedly obscure the outline of our argument.
% Therefore, as long as photon emissions are concerned,
% the main difference of this work from traditional OG models is that 
% synchro-curvature, ICS, and pair-creation processes near and outside the
% light cylinder are considered.

\subsection{Crab pulsar}
\label{sec:Crab}
\begin{figure}
\plotone{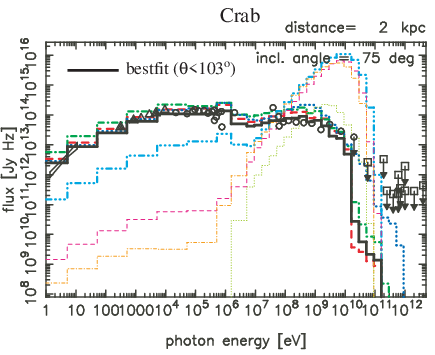}   % color (for electronic version)
% \plotone{f1_bw.eps}  % black and white (for printing version)
\caption{
Phase-averaged spectrum of the magnetospheric emissions from the Crab pulsar
for $\inc=75^\circ$, $\mu=4\times 10^{30}\mbox{\,G\,cm}^3$
and $h_{\rm m}=0.12$.
The thick dashed, dash-dotted, dotted and dash--triple-dotted
lines represent fluxes into 
$ 93^\circ<\theta<97^\circ$, $ 97^\circ<\theta<101^\circ$,
$101^\circ<\theta<105^\circ$,$105^\circ<\theta<109^\circ$;
the thin dashed, dash-dotted, dotted ones 
$109^\circ<\theta<113^\circ$,$113^\circ<\theta<117^\circ$,
$117^\circ<\theta<121^\circ$.
The thick solid line represents the averaged flux in
$75^\circ<\theta<103^\circ$.
%See Oke~(1969), Davidson et al.~(1982), Middleditch et al.~(1983), and
%Percival et al.~(1993), for IR, optical, UV data;
See Eikenberry et al. (1997) for IR-UV data;
%Harnden \& Seward~(1984), Pravdo \& Serlemitsos~(1981),
Knight~(1982), Weisskopf et al.~(2004), and
Mineo et al.~(2006) for X-rays;
Nolan et al.~(1993), Ulmer et al.~(1995),
Much et al.~(1995), Fierro et al.~(1998), Kuiper et al.~(2001)
for 10~MeV--20GeV;
Borione et al.~(1997), Tanimori et al.~(1998),
Hillas et al.~(1998), Lessard et al.~(2000), and
de Naurois et al.~(2002) for the upper limits above 50~GeV.
\label{fig:spc_crab}
}
\end{figure}

We present the results for the Crab pulsar 
in figure~\ref{fig:spc_crab},
adopting a magnetic inclination of $\inc=75^\circ$ and
a dipole moment of $\mu=4\times 10^{30}\mbox{\,G\,cm}^3$,
which is close to the value ($3.8\times 10^{30}\mbox{\,G\,cm}^3$)
deduced from the dipole radiation formula.
It follows that the observed pulsed spectrum from IR to VHE
can be reproduced,
provided that we observe the photons emitted into 
$75^\circ<\theta<103^\circ$,
where $\theta$ denotes the photon propagation direction
measured from the rotational axis.
Because of the aberration of light,
it is reasonable to suppose that photons emitted in a certain range of
$\theta$ comes into our line of sight
in an obliquely rotating three-dimensional magnetosphere
(e.g., RY95).
The flux rapidly decreases with decreasing $\theta$ for 
$75^\circ<\theta<93^\circ$,
because $\Ell$ is highly screened in the inner part of the gap.
Nevertheless, an average over 
$75^\circ < \theta < 103^\circ$,
which includes negligible flux between $75^\circ < \theta < 89^\circ$,
achieves the current objective,
because the spectral normalization can be fitted (within a factor of a few)
without changing the spectral shape,
by adjusting $h_{\rm m}$.

ICS takes place efficiently near and outside the light cylinder
(Aharonian \& Bogovalov 2003), 
because the magnetospheric IR photons, 
which are emitted along convex field lines, 
collide with the gap-accelerated positrons near the light cylinder, 
where the field lines are concave.
Most of such upscattered photons,
as well as some of the high-energy tail of the curvature component,
are absorbed by the $\gamma$-$\gamma$ collisions.
As a result, there is a gradual turnover around 10~GeV,
which forms a striking contrast with the steep turnover predicted 
in IG models due to magnetic pair creation.
The primary curvature component appears between 
100~MeV and 10~GeV,
while the secondary synchrotron component appears below a few MeV.
Between a few MeV and 100~MeV, the ICS component dominates,
because the secondary pairs that have been cooled down 
below a few hundred MeV
efficiently up-scatter magnetospheric UV and X-rays to lose energy.
%
% The strong synchrotron emission leads to a flat 
% spectrum in $10~\mbox{keV}<h\nu<\mbox{GeV}$.
Similar spectral shapes are obtained 
for different values of $\inc$, $\mu$, $h_{\rm m}$,
provided that the created current is super GJ,
by virtue of the negative feedback effects (H06).
% However, the flux changes within a factor of 3
% as a function of $h_{\rm m}$.
% Note that the spectral shape has a small dependence on $h_{\rm m}$
% once the created current becomes super-GJ.

\subsection{Vela pulsar}
\label{sec:Vela}
\begin{figure}
\plotone{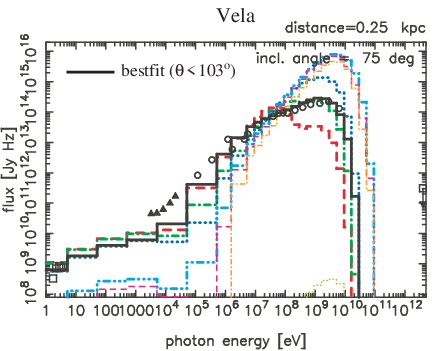}   % color (for electronic version)
\plotone{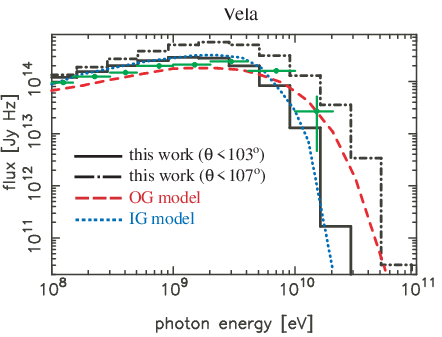}   % color (for electronic version)
% \plottwo{f2a.eps}{f2b.eps}    % color (for electronic version)
% \plottwo{f2a_bw.eps}{f2b_bw.eps} % black and white (for printing version)
\caption{
Phase-averaged Vela spectrum 
for $\mu=4\times 10^{30}\mbox{\,G\,cm}^3$,
$\inc=75^\circ$ and $h_{\rm m}=0.21$.
See Manchester et al. (1980) for optical data (open squares), 
Harding et al. (2002) for X-rays (filled triangles),
Kanbach et al. (1994), Ramanamurthy et al. (1995b), 
Fierro et al. (1998), Thompson et al. (1999)
for $\gamma$-rays (open circles).
{\it Left:} Same figure as figure~\ref{fig:spc_crab}.
{\it Right:} Close up above 100~MeV to
compare with IG (Harding \& Daugherty~1993) 
and traditional OG (Romani~1996) models.
The solid line is same as the left panel.
\label{fig:spc_vela}
}
\end{figure}

Next, we present the results for the Vela pulsar 
in figure~\ref{fig:spc_vela} (left).
Taking an angle average over $75^\circ<\theta<103^\circ$ (solid line),
we can reproduce the observed pulsed spectrum,
except for the RXTE results.
% If we set the azimuthal collision angle,
% which is assumed to vanish in the present paper,
% between the curvature $\gamma$-rays and the synchrotron X-rays,
% to be $\pi/16$~rad, we could reproduce the RXTE fluxes.
The primary curvature component appears between 
100~MeV and 10~GeV, 
while the secondary and tertiary synchrotron components appear below
100 MeV.
ICS is negligible for the Vela pulsar because of its weak magnetospheric
emission.
Similar spectral shapes are obtained for super-GJ solutions,
even though we have to adjust $h_{\rm m}$ to obtain an appropriate flux.

In the right panel, we compare the present results with IG (dotted)
and OG (dashed) models,
where the dash-dotted line denotes the averaged flux 
for $75^\circ <\theta< 107^\circ$.
It follows that the spectrum between
10 and 100 GeV crucially depends on the angles 
in which the photons that we observe are emitted.
Thus, to quantitatively predict the $\gamma$-ray emission
from the outer magnetosphere,
it is essential to examine the three-dimensional magnetic field structure
near and outside the light cylinder.
% Nevertheless, if we adopted finer momentum grids for the
% photon distribution function,
% we could, in principle, select the best-fit spectrum,
% which should be between the solid and dash-dotted lines.
% However, we cannot select the best-fit angle range 
% (e.g., $75^\circ-110^\circ$) in the present paper,
% becasue we are adopting $4^\circ$ as the increment of the photon 
% propagation directions.
% Before turning into a search of the best-fit angle range,
% we must extend the present analysis into the 3-D configuration space,
% correctly computing the magnetic field geometry near and outside the
% slight cylinder, which is out of the scope of this letter.

\subsection{PSR~B1951+32}
\label{sec:1951}
\begin{figure}
 \plotone{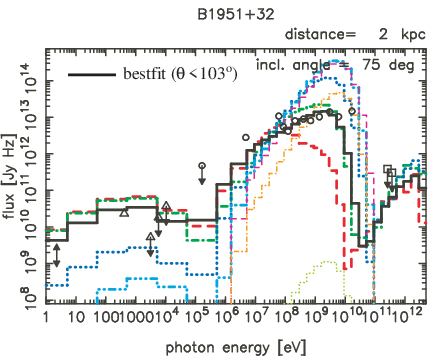}    % color (for electronic version)
 \plotone{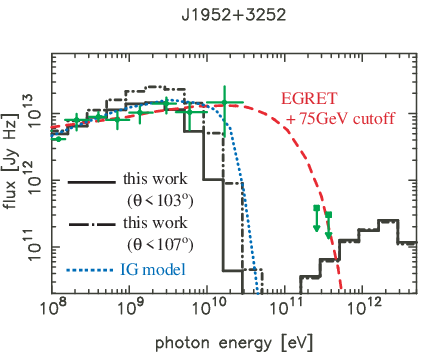}    % color (for electronic version)
% \plottwo{f3a.eps}{f3b.eps}    % color (for electronic version)
% \plottwo{f3a_bw.eps}{f3b_bw.eps} % black and white (for printing version)
\caption{
Phase-averaged spectrum of PSR~B1951+32 
for $\mu=2\times 10^{29}\mbox{\,G\,cm}^3$,
$\inc=75^\circ$ and $h_{\rm m}=0.39$.
Lines represent same quantities as figure~\ref{fig:spc_vela} 
unless notified.
See Clifton et al. (1988) for IR upper limits (open squares), 
Becker \& Tr\"umper (1996),
Safi-Harb et al. (1995) and Chang \& Ho (1997)
for X-ray data (open triangles),
Ramanamurthy et al. (1995a), for 100~MeV-15~GeV data (open circles),
% Albert et al. (2006) and  %%% If MAGIC data available, cite this ref.
Srinivasan et al.~(1997) for VHE upper limits.
\label{fig:spc_1951}
}
\end{figure}

Thirdly and finally, we present the spectrum of B1951+32
in figure~\ref{fig:spc_1951} (left).
It follows that the ROSAT and EGRET fluxes
are reproduced by taking the flux average in $75^\circ<\theta<103^\circ$
(solid line), 
except for 17~GeV flux, which was derived from only two photons.
Because of its weak magnetic field, less energetic synchrotron photons 
reduce the absorption of the ICS component.
The reduced absorption results in a small synchrotron flux,
which further reduces the absorption outside the light cylinder,
leading to a large, unabsorbed TeV fluxes.
% Here, the photon distribution function 
% is calculated between 0.005~eV and 5~TeV,
% instead of assuming it below 50~keV as in H06.
% Thus, the solved soft photon field inside the light cylinder
% determines the ICS flux near the light cylinder.
It also follows that the spectrum turns over at lower energy
than the IG model prediction 
(dotted curve in the right panel; Harding~2001).

It should be noted that 
the predicted ICS flux (above 100 GeV)
represents a kind of upper limits, 
because it is obtained by assuming that all the magnetospheric
synchrotron photons illuminate the equatorial region 
in which gap-accelerated positrons are migrating.
Some of such VHE photons materialize as energetic secondary pairs,
emitting the synchrotron component between a few keV and 100~MeV.
Between 100~MeV and 30~GeV, 
the primary curvature component dominates, 
which represents the absolute flux (instead of upper limits).
Some of such curvature photons have energies above 10~GeV
and are efficiently absorbed to materialize as less energetic pairs,
which emit synchrotron radiation below a few keV. 
Thus, the spectrum below a few keV also represents the absolute flux.

% In the right panel, we compare the present results
% with an IG prediction (dotted curve) and 
% an extrapolated EGRET spectrum (dashed curve).
% - - - - - - - - - - - - - - - - - - - - - - - - - - - - - - - - - 
% - - -  If MAGIC data available, select this block.
% It follows that both IG and current OG models are consistent with
% the very recent MAGIC upper limit (dash--triple-dotted curve)
% and that the current solution
% has a lower turnover energy even compared with the IG model.
% - - -  End of block selection.  - - - - - - - - - - - - - - - - - 
% - - - - - - - - - - - - - - - - - - - - - - - - - - - - - - - - - 
% If such photons do not efficiently illuminate this region,
% as suggested by a more or less straight field lines near the
% light cylinder for an oblique rotator (Spitkovski~2006),
% the ICS flux will decrease from the present estimate.
% Thus, to give a precise prediction of ICS fluxes, 
% we must extend the present analysis into three-dimensional 
% configuration space, 
% adopting a correct three-dimensional field line configuration
% inside and outside the light cyliner,
% which is out of the scope of this paper.

\
\section{Summary and Discussion}
\label{sec:summary}
To sum up, the self-consistent gap solutions
basically reproduce the observed power-law spectra below a few GeV
for the three pulsars examined.
The cut-off spectra between 10~GeV and 100~GeV
strongly reflect the three-dimensional magnetic field configuration
near the light cylinder;
thus, a discrimination between IG and OG models (e.g., using GLAST)
should be carefully carried out.
If pulsations are detected above 100~GeV,
it undoubtedly indicates that the photons are emitted via ICS
near the light cylinder, 
because VHE emissions cannot be expected in IG models.

% We adopted an average over emission angles 
% $75^\circ < \theta < 103^\circ$ for the solid curves in 
% figures~\ref{fig:spc_crab}, \ref{fig:spc_vela}, and \ref{fig:spc_1951}.
% Instead, we could take an average over
% $93^\circ < \theta < 103^\circ$,
% because the emission into $75^\circ < \theta < 93^\circ$
% is negligible for all the three pulsars, for $\inc=75^\circ$,
% owing to the small potential drop in the inner part of the gap.
% In this case, the flux increases 2.8 times.
% Nevertheless, the normalization of the spectra is adjustable by changing
% $h_{\rm m}$ and also susceptible to the toroidal structure of the gap, 
% which will be considered in subsequent papers. 
% Thus, we consider that an average over 
% $75^\circ < \theta < 103^\circ$ achieves the current objective.
 
Since our analysis is limited within the two-dimensional plane
formed by the magnetic field lines that thread the stellar surface
on the plane containing both the rotational and magnetic axes,
azimuthal structure is still unknown.
Thus, we cannot present pulse profiles, phase-resolved spectra,
or the polarization angle variations in this letter.
Since the gap is most active in the outer part of the magnetosphere
(unlike previous OG models, which adopt the vacuum solution
 of the Poisson equation and hence a uniform emissivity),
and since the photons will be emitted along the instantaneous
particle motion measured by a static observer
(unlike the treatment in the OG model of RY95,
 who assume a very strong aberration of light near the light cylinder),
it is possible that the predicted pulse profiles and so on are
quite different from previous OG models.
These topics will be discussed in the subsequent paper,
which deals with the three-dimensional gap structure 
near the light cylinder.

The pulsed TeV flux from PSR~B1951+32 can be predicted to be 
above $5 \times 10^{10}$~Jy\,Hz, 
provided that a certain fraction (more than $30\%$)
of the magnetospheric soft photons illuminate the equatorial region. 
However, if the poloidal field lines are more or less straight
near the light cylinder,
as demonstrated by Spitkovsky (2006, fig.~2) for an oblique rotator,
the equatorial region may not be efficiently illuminated.
In this case, the VHE flux will decrease from the current prediction.
There is room for further investigation
how to extend the present analysis into three dimensions,
and to combine it with time-dependence three-dimensional force-free
electrodynamics.
% to find a consistent magnetospheric current distribution
% between gap electrodynamics and force-free electrodynamics
% (i.e., smoothness of the field lines across the light cylinder),
% and to include plasma inertia, which will play an important role
% at the light cyliner where the coefficient of the highest-order
% derivative terms vanish in the equations of force-free electrodynamics.

\acknowledgments
The author is grateful to Drs.
N.~Otte, R.~Taam, J.~Takata 
for helpful suggestions.
% The bulk of this work was prepared while the author studied
This work is supported by the Theoretical Institute for
Advanced Research in Astrophysics (TIARA) operated under Academia Sinica
and the National Science Council Excellence Projects program in Taiwan
administered through grant number NSC 95-2752-M-007-006-PAE.

\clearpage

\clearpage

\clearpage

\end{document}